\documentclass[10pt, final, journal, letterpaper, twocolumn]{IEEEtran}

\makeatletter
\def\ps@headings{%
\def\@oddhead{\mbox{}\scriptsize\rightmark \hfil \thepage}%
\def\@evenhead{\scriptsize\thepage \hfil \leftmark\mbox{}}%
\def\@oddfoot{}%
\def\@evenfoot{}}
\makeatother 

\IEEEoverridecommandlockouts

\usepackage{amsfonts}
\usepackage[dvips]{graphicx}
\usepackage{caption}
\usepackage{subfigure}
\usepackage{times}
\usepackage{cite}
\usepackage{lettrine}
\usepackage{amsmath}
\usepackage{amsmath}
\allowdisplaybreaks[4]
\usepackage{array}
\usepackage{amssymb}

\usepackage{stfloats}
\usepackage{slashbox}
\usepackage{graphicx}
\usepackage{footnote}
\usepackage{booktabs}
\usepackage{array}
\usepackage{algorithmic}
\usepackage{algorithm}
\usepackage{subeqnarray}
\usepackage{cases}
\usepackage{threeparttable}
\usepackage{color}
\usepackage{multirow}
\usepackage{bm}
\usepackage{bbm}
\DeclareMathOperator*{\argmax}{argmax}

\newtheorem{proposition}{Proposition}[section]

\newtheorem{remark}{Remark}[section]

\begin{document}
\bibliographystyle{IEEEtran}

\title{Joint Uplink and Downlink Transmissions in User-Centric OFDMA Cloud-RAN}
\IEEEoverridecommandlockouts

\author{Zehong Lin,~\IEEEmembership{Student Member,~IEEE}, and Yuan Liu,~\IEEEmembership{Senior Member,~IEEE}


\thanks{Z. Lin was with the School of Electronic and Information Engineering,
South China University of Technology, Guangzhou, 510641, China. He is now with the Department of Information Engineering, The Chinese University of Hong Kong, Hong Kong (e-mail: zehonglin17@gmail.com).}
\thanks{Y. Liu is with the School of Electronic and Information Engineering, South China University of Technology, Guangzhou, 510641, China (email: eeyliu@scut.edu.cn).}
}

\maketitle

\vspace{-1.5cm}

\begin{abstract}
This paper studies joint uplink (UL) and downlink (DL) resource allocation in user-centric orthogonal frequency division multiple access (OFDMA) cloud radio access network (CRAN), where the users select the distributed remote radio heads (RRHs) to cooperatively serve their UL and DL transmissions over different subcarriers (SCs). The goal of this paper is to maximize the system throughput through jointly optimizing UL/DL scheduling, SC assignment, RRH grouping/clustering and power allocation under the maximum power and fronthaul capacity constraints. The problem is formulated as a mixed integer programming problem which is non-convex and NP-hard. We propose an efficient algorithm based on the Lagrange duality method to obtain an asymptotically optimal solution for this problem. A heuristic algorithm is further proposed to reduce the complexity. Simulation results illustrate that the proposed heuristic algorithm also has a close-to-optimal performance, and the proposed algorithms can considerably improve the system throughput compared to other benchmark schemes.
\end{abstract}


\section{Introduction}

To meet the rapidly increasing demand of mobile data in the wireless network, the upcoming fifth-generation (5G) wireless communication is expected to provide 1000 times higher throughput \cite{5G}. Network densification \cite{NetDen} is regarded as one of the promising ways to achieve this goal by increasing the density of the deployed base stations (BSs).

Cloud radio access network (CRAN) was first proposed by China Mobile \cite{CRAN1} and has been perceived as a promising candidate for the future 5G standard, reducing both the network capital expenditure (CAPEX) and operating expenditure (OPEX) \cite{CRAN2}. CRAN is a novel network architecture that enables centralized resource allocation at a baseband unit (BBU) pool by using coordinated multiple-point (CoMP) operation, and boosts both energy efficiency and spectral efficiency through network densification \cite{CRAN1, CRAN2, NetDen}. That is, the conventional BSs are replaced by cost-effective remote radio heads (RRHs) in CRAN. As a result, the RRHs can be deployed with a high density in the network due to their low operation cost and deployment cost. This will significantly reduce the distance between the RRHs and the users, and thus reduce the transmission power. In addition, the RRHs are coordinated by the BBU pool and cooperatively serve mobile users. A cluster of RRHs exchange information with the centralized BBU pool through high-speed wired/wireless fronthaul links \cite{CRAN3} and cooperatively forward the information between the BBU pool and users \cite{CRAN1}, while the task of baseband signal processing is left to the BBU pool. Therefore, significant performance gains can be achieved through joint network-level design and centralized signal processing. There are two clustering schemes for the RRHs to form clusters: cell-centric and user-centric. In the former scheme, each RRH selects users to serve and form a cluster or a cell, which is similar to conventional cellular systems and limits the performance of the cell-edge users. The latter scheme allows each user to associate with a set of RRHs and form a cluster, and thus it eliminates cell-edge users. In general, the user-centric scheme outperforms the cell-centric scheme. In this paper, we consider user-centric CRAN.

Recently, several works have investigated various resource allocation problems in CRAN \cite{CRAN4, CRAN5, CRAN6, clustering, CRAN7, CRAN8, CRAN9, CRAN10}. For instance, the authors in \cite{CRAN4} studied the problem of joint RRH selection and power minimization through coordinated beamforming subject to users' quality-of-service (QoS) requirements. The work \cite{CRAN5} studied a similar problem and particularly considered joint uplink (UL) and downlink (DL) user association and beamforming design. In \cite{CRAN6}, the authors studied the problem of joint precoding and RRH selection to minimize the network power consumption. The authors in \cite{clustering} considered sparse beamforming based clustering to maximize the downlink weighted sum rate. Note that the works \cite{CRAN4, CRAN5, CRAN6, clustering} did not take fronthaul constraints into account and considered narrow-band transmission. In \cite{CRAN7}, the authors considered uniform scalar quantization in an orthogonal frequency division multiple access (OFDMA) CRAN and studied the throughput maximization problem with joint fronthaul allocation and power control, where subcarrier (SC) allocation was not considered. The works \cite{CRAN8, CRAN9, CRAN10} studied resource allocation in OFDMA-based CRAN.

Note that in the above works \cite{CRAN4, CRAN6, clustering, CRAN7, CRAN8, CRAN9, CRAN10}, only either UL or DL was considered. However, it is necessary to consider both UL and DL transmissions while optimizing resource allocation. Although {5G} systems are dominated by DL traffic, UL transmission is becoming more and more important due to the increasing high-demand UL applications, such as high definition video calling and online gaming. The optimal DL transmission may not be optimal for UL because of the asymmetric traffic between DL and UL. Therefore, UL and DL should be designed jointly in practice. If joint UL/DL transmission is considered, a user's UL and DL can associate with different BSs because the channel conditions, transmit power, and QoS requirements of UL and DL for the same user could be largely different in cellular systems (especially in small cells) \cite{DUDe3}. This is known as UL/DL decoupling. The problem becomes more complicated and challenging in CRAN because the UL and DL of a user can associate with different groups of RRHs if UL/DL decoupling is considered, i.e., a user should select different groups of RRHs to form CoMP for its UL and DL from the user-centric perspective. Although \cite{CRAN5} considered both DL and UL, it considered narrow-band transmission where all the RRHs and users operate on the same spectrum. Note that OFDMA-based multiuser transmission is more preferable for high-speed demanding applications. More importantly, in OFDMA systems, the parallel transmission structure of OFDMA channels allows UL and DL traffic to be scheduled across different SCs \cite{FDD1, FDD2, FDD3, DRA1, DRA2, DRA3}, which opens a new dimension for flexible UL/DL resource allocation. The flexible DL and UL resource allocation has been investigated in traditional cellular networks \cite{DRA1, DRA2, DRA3}. In particular, the authors in \cite{DRA1} investigated the dynamic time division duplex (TDD) resource allocation in both homogeneous and heterogeneous networks. Considering the effect of adjacent channel emissions, \cite{DRA2} studied the coexistence of flexible FDD where the UL band can be utilized for DL. In \cite{DRA3}, the authors studied flexible FDD with power control and network-based interference cancellation to support dynamic asymmetric DL/UL traffic. However, this issue has not been considered in CRAN to date. Furthermore, if OFDMA is adopted, the network resources, like SCs, UL/DL traffic, and RRH clustering based association are affected by each other and thus should be optimized jointly. The above considerations motivate our work.

\begin{figure}[t]
\begin{centering}
\includegraphics[scale=0.34]{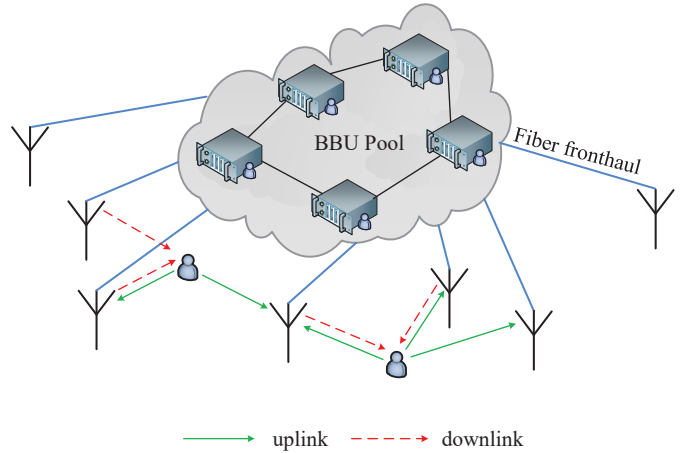}
\vspace{-0.1cm}
 \caption{System model of the considered user-centric OFDMA-based CRAN.}\label{fig:system}
\end{centering}
\vspace{-0.1cm}
\end{figure}

In this paper, we design joint UL and DL resource allocation in a user-centric OFDMA-based CRAN as shown in Fig. \ref{fig:system}, where multiple users associate with RRHs for both UL and DL transmissions using OFDMA. The main contributions of this paper are summarized as follows:
\begin{itemize}
    \item We propose a new user-centric CRAN scheme where UL and DL take place at the same time but over different orthogonal SCs. Specifically, each SC can be assigned to either UL or DL of a user for decoupled bidirectional transmission, and each user is allowed to associate with a different group of RRHs over each SC, i.e., UL/DL decoupling as well as UL/DL CoMP are carried out via the user-centric basis. The proposed scheme can adapt to different UL/DL traffic demands and optimize the resource allocation flexibly.

    \item Based on the proposed user-centric CRAN scheme, we study a joint resource allocation problem of UL/DL scheduling, SC assignment, RRH clustering, and power allocation to maximize the system throughput. To our best knowledge, this paper is the first attempt to study joint UL and DL resource allocation problem in an OFDMA-based CRAN. The user-centric CoMP over different SCs for joint UL and DL transmissions make the problem and solution significantly different and complicated.

    \item The formulated problem is a mixed integer programming problem, which is non-convex and NP-hard. We propose an efficient algorithm based on Lagrange duality method, which solves the problem asymptotically optimally when the number of SCs is large. Moreover, we propose a heuristic algorithm to simplify the RRH selection, which provides a good tradeoff between complexity and performance.
\end{itemize}

The rest of the paper is organized as follows. In Section II, the system model and problem formulation are presented. The proposed asymptotically optimal and heuristic algorithms are given in Section III and Section IV, respectively. Section V discusses the case of time division duplex (TDD) mode and Section VI provides simulation results and analysis. Finally, conclusions are made in Section VII.

\section{System Model and Problem Formulation}

In this section, we first describe the system model of the considered user-centric CRAN. Then we formulate the throughput maximization problem.

\subsection{System Model}

We consider an OFDMA-based CRAN where exists $K$ single-antenna users and $M$ single-antenna RRHs, as shown in Fig. \ref{fig:system}. All nodes are half-duplex due to practical consideration. Half-duplex is more practical and cost-effective since transmitting and receiving happen in different time or frequency without self-interference, and thus it is easy to decode signals. It is assumed that each RRH connects to the BBU pool for both UL and DL information exchange via a fiber fronthaul link, and each RRH communicates with users over wireless channels. The BBU pool is able to conduct joint signal processing and centralized resource allocation. The spectrum is equally divided into $N$ SCs, and each SC can be assigned to at most one user to avoid inter-user interference. Let $\mathcal{K}$ = $\{ 1, \cdots, K\}$ denote the set of users, $\mathcal{M}$ = $\{ 1, \cdots, M\}$ the set of RRHs, and $\mathcal{N} = \{1, \cdots, N\}$ the set of SCs. We assume that the additive white Gaussian noises (AWGN) at all nodes are independent circular symmetric complex Gaussian random variables with zero mean and variance $\sigma^2$. The channel fading is modeled by large-scale path loss fading and small-scale fading. The transmission is divided into successive slots, where the channel fading remains unchanged within each slot but varies independently from one slot to another, i.e., the channel impulse response can be treated as invariant within each slot. The channel state information (CSI) can be estimated at each user or RRH. Specifically, a user (RRH) transmits a pilot and each associated RRH (user) receives the pilot and estimates CSI, then the estimated CSI is fed back to the central controller, i.e., the BBU pool, for signal processing. Therefore, we assume that perfect CSI is available for resource processing at the BBU pool.

Here, we assume that UL and DL occur at the same time but over different SCs \cite{FDD1,FDD2,FDD3}, i.e., the \emph{flexible} frequency division duplex (FDD) mode \cite{DRA2, DRA3}. To avoid inter-link interference, each SC can be assigned to either UL or DL of a user. To this end, let the binary variable $y_{n}$ indicate the UL/DL scheduling on SC $n$, i.e.,
\begin{align}
    y_{n} =
    \begin{cases}
    1,   &\mbox{if SC $n$ is assigned to UL},\\
    0,   &\mbox{if SC $n$ is assigned to DL}.
    \end{cases}
\end{align}

Each user can select a subset of RRHs to perform CoMP \cite{CoMP} on each SC for either UL or DL. The joint SC assignment and RRH selection on SC $n$ is indicated by a binary decision variable $x_{k, m, n}$, and we have
\begin{align}
    x_{k, m, n} =
    \begin{cases}
    1,   &\mbox{if user $k$ selects RRH $m$ on SC $n$ },\\
    0,   &\mbox{otherwise}.
    \end{cases}
\end{align}
Define $\mathbf{X}_{n} = [x_{1,1,n} \cdots x_{1,M,n}; \cdots; x_{K,1,n} \cdots x_{K,M,n}] \in \{0, 1\}^{K \times M}$ as the joint SC assignment and RRH selection matrix on SC $n$. Denote $k_{n}$ as the user assigned to SC $n$ and the corresponding subset of serving RRHs as
\begin{align}
    \mathcal{M}_{n} = \{m \in \mathcal{M} | x_{k_{n}, m, n} = 1\}, n \in \mathcal{N}.
\end{align}
Thus, the RRHs in $\mathcal{M}_{n}$ serve $k_{n}$ for UL on SC $n$ if $y_{n} = 1$, otherwise for DL if $y_{n} = 0$. In the following, we present the signal models for UL and DL in detail, respectively.

\begin{remark} \label{remark_1}
In practical FDD systems, UL and DL occur over different frequency bands and need guard band to avoid interference. Our system model and algorithms can be applicable to such a practical case: The whole spectrum is divided into two bands, with one for UL and the other for DL. There exists a guard band between the UL and DL bands for avoiding interference. Both the UL and DL bands are equally divided into SCs for dedicated UL and DL transmissions, respectively. Although the UL and DL bands are fixed in practical FDD systems (i.e., $y_n$'s are fixed), the SCs in both UL and DL bands need to be assigned to different users and RRHs (i.e., $x_{k,m,n}$ should be optimized). Thus the proposed \emph{flexible FDD} framework can accommodate the \emph{practical} FDD system by fixing $y_n$.
\end{remark}

In UL, we use $h_{k, m, n}^{u}$ to denote the complex channel coefficient from user $k$ to RRH $m$ on SC $n$. Let $\mathbf{p}_{n}^u = [p_{1, n}^u, \cdots, p_{K, n}^u]^\mathrm{T} \in \mathbb{R}_{+}^{K \times 1}$ denote the transmit power vector for the $K$ users on SC $n$, and $p_{k, n}^u$ is the transmit power of user $k$ on SC $n$. Thus, the signal received at RRH $m \in \mathcal{M}_{n}$ on SC $n$ can be represented as
\begin{align}
S_{m, n} = s^{u}_{k, n} \sqrt{x_{k, m, n} p_{k, n}^u|h_{k, m, n}^u|^{2}} + n_{m},
\end{align}
where $s^{u}_{k, n}$ denotes the information symbols transmitted by user $k$ on SC $n$, and $n_{m}$ is the AWGN at RRH $m$.

Since the fronthaul capacity is limited in practice, we assume that for UL the RRHs first quantize the received signals and then forward to the BBU pool over the fronthaul. Specifically, each RRH performs uniform scalar quantization \cite{CRAN7, CRAN9} independently on each SC. The quantized signal of $S_{m, n}$ is given by
\begin{align}
\tilde{S}_{m, n} = S_{m, n} + e_{m, n},
\end{align}
where $e_{m, n}$ is the quantization error with zero mean and variance $q_{m, n}$. As the uniform scalar quantization is performed by each RRH independently on each SC, the errors $e_{m, n}$'s are independent over $m$ and $n$. We assume that the resolution of the uniform scalar quantization is fixed and the same on all SCs for all RRHs, and we use $\beta$ to denote it. Therefore, the in-phase (I) and quadrature (Q) parts of the received complex signal $S_{m, n}$ can be represented by $\beta$ bits, respectively. Then the variance $q_{m, n}$ of the quantization error $e_{m, n}$ can be expressed as \cite{CRAN7, CRAN9}
\begin{align}
    q_{m, n} = 3 (p_{k, n}^u|h_{k, m, n}^u|^{2} + \sigma^2)2^{-2 \beta}.
\end{align}
After quantization, each RRH encodes the quantized values into digital codewords and transmits them to the BBU pool via the fronthaul link.

After receiving the digital codewords, the BBU pool recovers the quantized signals from the digital codewords and performs maximal-ratio combining (MRC) technique over the quantized signals on SC $n$ to jointly decode user $k$'s information. Thus, the UL rate of user $k$ on SC $n$ can be expressed as:
\begin{align}
R_{k,n}^{u} = \log_{2}\bigg(1 + \sum\limits_{m \in \mathcal{M}} \frac{x_{k, m, n} p_{k, n}^u |h_{k, m, n}^u|^{2}}{\sigma^2 + 3 (p_{k, n}^u|h_{k, m, n}^u|^{2} + \sigma^2)2^{-2 \beta}} \bigg). \label{eqn:ku}
\end{align}

In DL, we use $h_{k, m, n}^d$ to denote the complex channel coefficient from RRH $m$ to user $k$ on SC $n$. Let $\mathbf{p}_{n}^d = [p_{1, n}^d, \cdots, p_{M, n}^d]^\mathrm{T} \in \mathbb{R}_{+}^{M \times 1}$ denote the transmit power vector for the $M$ RRHs on SC $n$, and $p_{m, n}^d$ denote the transmit power of RRH $m$ on SC $n$. Note that the DL over each SC is actually a multiple-input single-output (MISO) channel \cite{MISO}. Thus, the signal received at user $k$ on SC $n$ can be represented as
\begin{align}
S_{k, n} = s^{d}_{k, n} \sum\limits_{m \in \mathcal{M}} \sqrt{x_{k, m, n} p_{m, n}^d|h_{k, m, n}^d|^{2}} + n_{k},
\end{align}
where $s^{d}_{k, n}$ denotes the information symbols for user $k$ on SC $n$, and $n_{k}$ is the AWGN at user $k$. Thus, we can obtain the DL rate of user $k$ on SC $n$ as:
\begin{align}
R_{k, n}^{d} = \log_{2}\Bigg(1 + \frac{1}{\sigma^2} \bigg(\sum\limits_{m \in \mathcal{M}} \sqrt{x_{k, m, n} p_{m, n}^d |h_{k, m, n}^d|^{2}} \bigg)^{2} \Bigg). \label{eqn:kd}
\end{align}

\subsection{Problem formulation}

In this paper, our objective is to maximize the system throughput by jointly optimizing the UL/DL scheduling, SC assignment, RRH selection and power allocation. Therefore, the problem can be mathematically formulated as:
\begin{subequations}
\begin{align}
\max_{\{\mathbf{X}, \mathbf{y}, \mathbf{P}_{u},\mathbf{P}_{r}\}}~&R_{total}= \sum\limits_{n \in \mathcal{N}}\sum\limits_{k \in \mathcal{K}}\Big[w_{k} y_{n} R_{k,n}^{u} + (1 - y_{n}) R_{k,n}^{d}\Big] \label{problem}  \\
{\rm s.t.}~&\sum\limits_{n \in \mathcal{N}} p_{k, n}^u \leq P_{k}^u, \forall k \in \mathcal{K}, \label{cons1} \\
  &\sum\limits_{n \in \mathcal{N}} p_{m, n}^d \leq P_{m}^d, \forall m \in \mathcal{M}, \label{cons2} \\
  &\sum\limits_{n \in \mathcal{N}} \sum\limits_{k \in \mathcal{K}} x_{k, m, n}  \leq C_{m},\forall m \in \mathcal{M}, \label{cons3} \\
  &\sum\limits_{k \in \mathcal{K}} x_{k, m, n}  \leq 1,\forall n \in \mathcal{N}, m \in \mathcal{M}, \label{cons4}   \\
  &x_{k, m, n} \in \{0, 1\}, \forall k \in \mathcal{K}, m \in \mathcal{M}, n \in \mathcal{N}, \label{cons5}   \\
  &y_{n} \in \{0, 1\}, \forall n \in \mathcal{N}, \label{cons6}
\end{align}
\end{subequations}
where $\mathbf{X} \triangleq \{\mathbf{X}_{n}\}$, $\mathbf{y} \triangleq \{y_{n}\}$, $\mathbf{P}_{u} \triangleq \{\mathbf{p}_{n}^u\}$ and $\mathbf{P}_{r} \triangleq \{\mathbf{p}_{n}^d\}$. In \eqref{problem}, weight $w_{k}$ accounts for asymmetric traffic of UL and DL for user $k$. In \eqref{cons1} and \eqref{cons2}, $P_{k}^u$ and $P_{m}^d$ denote the maximum power budget at user $k$ and RRH $m$, respectively. The constraint \eqref{cons3} indicates that RRH $m$ can occupy at most $C_{m}$ SCs, which guarantees a certain degree of fronthaul capacity constraints of the RRHs. \eqref{cons4} constrains that each SC can be allocated to at most one user, while the user can associate with multiple RRHs.

\begin{remark}
The fronthaul capacity is usually captured in terms of rate. Since the bandwidth of each SC is fixed, the channel capacity on each SC is also fixed. Therefore, the number of SCs cannot exactly represent the fronthaul capacity, but at least to some degree. Thus, in this paper, we use the number of SCs to simplify the fronthaul constraint of each RRH. Note that we can also use the rate to represent the fronthaul capacity, but this will greatly complicate the problem.
\end{remark}

Problem \eqref{problem} is non-convex due to integer constraints \eqref{cons3}-\eqref{cons6}. Note that if without the power allocation, the reduced problem is a joint SC assignment and RRH selection problem, which is combinatorial and NP-hard (please see the proof in Appendix \ref{appendicesA}). As a result, the considered Problem \eqref{problem} is more complex and also NP-hard. The exhaustive search requires to search $\big(2^{M + 1} K\big)^{N}$ probabilities for finding the optimal solution, which is complexity-prohibitive when the values of $N$ and/or $M$ are large.

\section{Asymptotically Optimal Solution}
Although Problem \eqref{problem} is NP-hard due to the coupled continuous and discrete variables, the works \cite{dual1, dual2, dual3} show that the so-called ``time-sharing" condition is satisfied in OFDMA systems regardless of the non-convexity of the original problem.

\begin{proposition}\label{proposition1}
Problem \eqref{problem} satisfies the ``time-sharing" condition as $N \rightarrow \infty$.
\end{proposition}

\begin{IEEEproof}
Please refer to Appendix \ref{appendicesB}.
\end{IEEEproof}

Proposition \ref{proposition1} implies that Problem \eqref{problem} can be solved by the Lagrange duality method asymptotically optimally since $N$ is typically large in practice.

\subsection{Dual Decomposition}
Let $\lambda_{k}$ denote the Lagrange multiplier (dual variable) corresponding to the constraints in \eqref{cons1} and $\bm{\lambda} \triangleq \{\lambda_{k}\}$. Also let $\mu_{m}$ and $\nu_{m} $ denote the Lagrange multipliers corresponding to the constraints in \eqref{cons2} and \eqref{cons3}, respectively, and $\bm{\mu} \triangleq \{\mu_{m}\}$ and $\bm{\nu} \triangleq \{\nu_{m}\}$. The Lagrangian function of Problem \eqref{problem} is given by equation \eqref{lagrangian function}, shown at the top of the next page.
\begin{figure*}[th]
\begin{align}
    \mathcal{L}(\mathbf{X}, \mathbf{y}, \mathbf{P}_{u},\mathbf{P}_{r},\bm{\lambda},\bm{\mu}, \bm{\nu})
    &= \sum\limits_{n \in \mathcal{N}}\sum\limits_{k \in \mathcal{K}}\Big[w_{k} y_{n} R_{k,n}^{u} + (1 - y_{n}) R_{k,n}^{d}\Big] - \sum\limits_{k \in \mathcal{K}} \lambda_{k} \bigg(\sum\limits_{n \in \mathcal{N}} p_{k, n}^u - P_{k}^u \bigg)  \nonumber \\
    &~~~ - \sum\limits_{m \in \mathcal{M}} \mu_{m} \bigg(\sum\limits_{n \in \mathcal{N}} p_{m, n}^d - P_{m}^d \bigg) - \sum\limits_{m \in \mathcal{M}} \nu_{m} \bigg(\sum\limits_{n \in \mathcal{N}}\sum\limits_{k \in \mathcal{K}} x_{k, m, n} - C_{m}\bigg) \nonumber \\
    &= \sum\limits_{n \in \mathcal{N}}\sum\limits_{k \in \mathcal{K}}\Big[w_{k} y_{n} R_{k,n}^{u} + (1 - y_{n}) R_{k,n}^{d}\Big] - \sum\limits_{k \in \mathcal{K}} \lambda_{k} \sum\limits_{n \in \mathcal{N}} p_{k,n}^u \nonumber \\
    &~~~ - \sum\limits_{m \in \mathcal{M}} \bigg(\mu_{m} \sum\limits_{n \in \mathcal{N}} p_{m, n}^d  + \nu_{m} \sum\limits_{n \in \mathcal{N}}\sum\limits_{k \in \mathcal{K}} x_{k, m, n}\bigg) + \sum\limits_{k \in \mathcal{K}} \lambda_{k}P_{k}^u + \sum\limits_{m \in \mathcal{M}} \bigg(\mu_{m}P_{m}^d + \nu_{m}C_{m} \bigg). \label{lagrangian function}
    \end{align}
    \hrulefill
\end{figure*}

Define $\mathcal{S}$ as the set of $\{\mathbf{X}, \mathbf{y}, \mathbf{P}_{u},\mathbf{P}_{r}\}$ satisfying the primary constraints, then the Lagrange dual function of Problem \eqref{problem} can be expressed as
\begin{align}
g(\bm{\lambda},\bm{\mu}, \bm{\nu})= \max_{\{\mathbf{X}, \mathbf{y}, \mathbf{P}_{u},\mathbf{P}_{r}\} \in \mathcal{S}} \mathcal{L}(\mathbf{X}, \mathbf{y}, \mathbf{P}_{u},\mathbf{P}_{r},\bm{\lambda},\bm{\mu}, \bm{\nu}). \label{dual function}
\end{align}

From \eqref{lagrangian function}, we can observe that the maximization in \eqref{dual function} can be decomposed into $N$ independent subproblems, and each subproblem corresponds to a particular SC. Therefore, we can rewrite the Lagrangian function as
\begin{align}
&\mathcal{L}(\mathbf{X}, \mathbf{y}, \mathbf{P}_{u},\mathbf{P}_{r},\bm{\lambda},\bm{\mu}, \bm{\nu}) \nonumber \\
&= \sum\limits_{n \in \mathcal{N}} \mathcal{L}_{n}(\mathbf{X}_{n}, y_{n}, \mathbf{p}_{n}^u,\mathbf{p}_{n}^d,\bm{\lambda},\bm{\mu}, \bm{\nu}) + \sum\limits_{k \in \mathcal{K}} \lambda_{k}P_{k}^u \nonumber \\
&~~~~  + \sum\limits_{m \in \mathcal{M}} \bigg(\mu_{m}P_{m}^d + \nu_{m}C_{m} \bigg), \label{eqn:L}
\end{align}
where
\begin{align}
&\mathcal{L}_{n}(\mathbf{X}_{n}, y_{n}, \mathbf{p}_{n}^u,\mathbf{p}_{n}^d,\bm{\lambda},\bm{\mu}, \bm{\nu}) \nonumber \\
&\triangleq \sum\limits_{k \in \mathcal{K}}\Big[w_{k} y_{n} R_{k,n}^{u} + (1 - y_{n}) R_{k,n}^{d}\Big] - \sum\limits_{k \in \mathcal{K}} \lambda_{k} p_{k,n}^u \nonumber \\
&~~~ - \sum\limits_{m \in \mathcal{M}} \bigg(\mu_{m} p_{m, n}^d  + \nu_{m} \sum\limits_{k \in \mathcal{K}} x_{k, m, n}\bigg).
\end{align}
As the last two terms in \eqref{eqn:L} are constants, maximizing $\mathcal{L}$ is equivalent to maximizing $\mathcal{L}_{n}$ on each SC $n$. The subproblem on SC $n$ can be further decoupled to two subproblems, with one for UL and the other for DL, i.e.,
\begin{align}
\max_{\{\mathbf{X}_{n}, y_{n}, \mathbf{p}_{n}^u,\mathbf{p}_{n}^d\}} \mathcal{L}_{n} = y_{n} \mathcal{L}_{n}^{u} + (1 - y_{n}) \mathcal{L}_{n}^{d}, \label{subproblem}
\end{align}
where
\begin{align}
\mathcal{L}_{n}^{u} = \sum\limits_{k \in \mathcal{K}}\Big(w_{k} R_{k,n}^{u}  - \lambda_{k} p_{k,n}^u - \sum\limits_{m \in \mathcal{M}} \nu_{m} x_{k, m, n}\Big),
\end{align}
and
\begin{align}
\mathcal{L}_{n}^{d} = \sum\limits_{k \in \mathcal{K}} R_{k,n}^{d} - \sum\limits_{m \in \mathcal{M}} \Big(\mu_{m} p_{m, n}^d + \nu_{m} \sum\limits_{k \in \mathcal{K}} x_{k, m, n}\Big) .
\end{align}
Therefore, in the following, we only focus on solving the subproblem \eqref{subproblem}.
\subsection{Optimizing $\{\mathbf{X}, \mathbf{y}, \mathbf{P}_{u},\mathbf{P}_{r}\}$ for Given $\{\bm{\lambda},\bm{\mu}, \bm{\nu}\}$}

For given $\{\bm{\lambda},\bm{\mu}, \bm{\nu}\}$, we can solve each subproblem \eqref{subproblem} to obtain the optimal $\{\mathbf{X}_{n}, y_{n}, \mathbf{p}_{n}^u,\mathbf{p}_{n}^d\}$ as follows.

\subsubsection{Maximizing Lagrangian over $\big\{\mathbf{p}_{n}^u,\mathbf{p}_{n}^d\big\}$} \label{p_sec}
For given $\{y_{n}, \mathbf{X}_{n}\}$, we now derive the optimal UL and DL power allocations $\mathbf{p}_{n}^{u*}$ and $\mathbf{p}_{n}^{d*}$ in the following.

With $y_n = 1$ and given $\mathbf{X}_{n}$, subproblem \eqref{subproblem} reduces to
\begin{align}
\max_{p_{k_n, n}^u \geq 0} {\mathcal{L}_n^{u}}' = w_{k_n} R_{k_n, n}^{u}  - \lambda_{k_n} p_{k_n, n}^u. \label{pa_ul}
\end{align}
Besides, we have $\mathbf{p}_{n}^{d*} = \{\mathbf{0}\}^{M \times 1}$ and $p_{k, n}^u = 0$ if $k \neq k_{n}$. We can easily prove that the objective of problem \eqref{pa_ul}, i.e., ${\mathcal{L}_n^{u}}'$, is concave in $p_{k_n, n}^u$, thus problem \eqref{pa_ul} is convex. Therefore, the optimal power allocation $p_{k_n, n}^{u*}$ can be obtained efficiently by one-dimensional line search \cite{CRAN9}. Note that if there is only one RRH in the UL, the optimal power allocation $p_{k_n, n}^{u*}$ can be obtained in the closed-form as given by the following proposition.

\begin{proposition}\label{proposition2}
Let $m_n$ be the single RRH in the UL. The optimal power allocation $p_{k_n, n}^{u*}$ for problem \eqref{pa_ul} is given by
\begin{align}
    p_{k_n, n}^{u*} = &\frac{\sigma^2}{2 \eta |h_{k_n, m_n, n}^u|^2} \mathcal{A}_{k_{n}, m_n, n} \label{p_ul}
\end{align}
where
\begin{align}
    \mathcal{A}_{k_{n}, m_n, n} =  \Bigg[\sqrt{1 + \frac{4w_{k_n}\eta |h_{k_n, m_n, n}^u|^2 }{\lambda_{k_n}\sigma^2 \ln2}} - (1 + 2\eta) \Bigg]^{+}, \label{An}
\end{align}
and $\eta = 3 / 2^{2\beta}$, $[\cdot]^{+}= \max\{\cdot, 0\}$.
\end{proposition}

\begin{IEEEproof}
\begin{figure*}[tb]
\begin{align}
    \frac{\partial {\mathcal{L}_{n}^u}'}{\partial p_{k_n, n}^u} = w_{k_n} \frac{|h_{k_n, m_n, n}^u|^{2}\sigma^2 (1 + 3 \cdot 2^{-2 \beta})}{\ln 2 \bigg(1 + \frac{|h_{k_n, m_n, n}^u|^{2} p_{k_n, n}^u}{\sigma^2 + 3(|h_{k_n, m_n, n}^u|^{2} p_{k_n, n}^u + \sigma^2)2^{-2 \beta}}\bigg) \Big(\sigma^2 + 3(|h_{k_n, m_n, n}^u|^{2} p_{k_n, n}^u + \sigma^2)2^{-2 \beta}\Big)^2} - \lambda_{k_n}  = 0 \label{Ln_ul}
    \end{align}
    \hrulefill
\end{figure*}
By applying the optimality Karush-Kuhn-Tucker (KKT) conditions \cite{ConvexOptimization} with respect to $p_{k_n, n}^u$, we can obtain equation \eqref{Ln_ul} shown at the top of the next page, which can be rearranged as
\begin{align}
    - \frac{\lambda_k \eta |h_{k_n, m_n, n}^u|^4}{\sigma^4} {p_{k_n, n}^u}^2 &- \frac{\lambda_k (1 + 2\eta) |h_{k_n, m_n, n}^u|^2}{\sigma^2} p_{k_n, n}^u \nonumber \\
    &- \lambda_k(1 + \eta) + \frac{w_k|h_{k_n, m_n, n}^u|^2}{\sigma^2 \ln2} = 0. \label{Ln_ul2}
\end{align}
The roots of the quadratic equation \eqref{Ln_ul2} are given by
\begin{align}
     r_1 = \frac{-\sigma^2}{2 \eta |h_{k_n, m_n, n}^u|^2} \bigg(\sqrt{1 + \frac{4w_{k_n}\eta |h_{k_n, m_n, n}^u|^2 }{\lambda_{k_n}\sigma^2 \ln2}} + (1 + 2\eta) \bigg),
\end{align}
and
\begin{align}
     r_2 = \frac{\sigma^2}{2 \eta |h_{k_n, m_n, n}^u|^2} \bigg(\sqrt{1 + \frac{4w_{k_n}\eta |h_{k_n, m_n, n}^u|^2 }{\lambda_{k_n}\sigma^2 \ln2}} - (1 + 2\eta) \bigg).
\end{align}

It is obvious that $p_{k_n, n}^{u*} = r_1 < 0$ is not available since we require that $p_{k_n, n}^{u*} \geq 0$. To ensure that $p_{k_n, n}^{u*} = r_2 \geq 0$, we must have
\begin{align}
    \sqrt{1 + \frac{4w_{k_n}\eta |h_{k_n, m_n, n}^u|^2 }{\lambda_{k_n}\sigma^2 \ln2}} - (1 + 2\eta) \geq 0.
\end{align}
Therefore, the optimal power allocation $p_{k_n, n}^{u*}$ is obtained as given in \eqref{p_ul}.
\end{IEEEproof}

With $y_n = 0$ and given $\mathbf{X}_{n}$, we have $\mathbf{p}_{n}^{u*} = \{\mathbf{0}\}^{K \times 1}$ and subproblem \eqref{subproblem} reduces to
\begin{align}
\max_{\mathbf{p}_{n}^d} {\mathcal{L}_n^{d}}' = R_{k_n, n}^{d} - \sum\limits_{m \in \mathcal{M}_n} \mu_{m} p_{m, n}^d. \label{pa_dl}
\end{align}
It is obvious that the objective of problem \eqref{pa_dl} is concave in $\mathbf{p}_{n}^d$. The optimal power allocation $\mathbf{p}_{n}^d$ can be obtained in the closed-form as given by the following proposition.
\begin{proposition}\label{proposition3}
The optimal DL power allocations for problem \eqref{pa_dl} is given by
\begin{align}
    p_{m, n}^{d*} =
    \begin{cases}
    0,   &\mbox{$m \notin \mathcal{M}_n$}\\
    \frac{|h_{k_n, m, n}^d|^{2}}{\sigma^2 \mu_{m}^{2} \big(\mathcal{B}_{k_{n},n}\big)^{2}}\bigg[\frac{1}{\ln 2} \mathcal{B}_{k_{n},n} - 1 \bigg]^{+},   &\mbox{$m \in \mathcal{M}_n$},
    \end{cases} \label{P_r1}
\end{align}
where
\begin{align}
    \mathcal{B}_{k_{n},n} = \sum\limits_{m \in \mathcal{M}_{n}} \frac{|h_{k_n, m, n}^d|^{2}}{\sigma^2 \mu_{m}}, \label{Bn}
\end{align}
and $[\cdot]^{+}= \max\{\cdot, 0\}$.
\end{proposition}
\begin{IEEEproof}
By applying the optimality KKT conditions with respect to $\mathbf{p}_{n}^d$, we can obtain that
\begin{align}
    \frac{\partial {\mathcal{L}_n^{d}}'}{\partial p_{m, n}^d}
    &= \frac{\Big(\frac{1}{\sigma}\sum\limits_{j \in \mathcal{M}_{n}}\sqrt{p_{j, n}^d |h_{k_{n}, j, n}^d|^2}\Big) \sqrt{\frac{|h_{k_{n}, m, n}^d|^2}{\sigma^2}}}{\ln 2 \bigg[1 + \Big(\frac{1}{\sigma}\sum\limits_{j \in \mathcal{M}_{n}}\sqrt{p_{j, n}^d |h_{k_{n}, j, n}^d|^2}\Big)^{2}\bigg]\sqrt{p_{m, n}^d}} - \mu_{m} \nonumber \\
    &= 0, \label{Pr}
\end{align}
$\forall m \in \mathcal{M}_{n}, y_{n} = 0$.
With the optimal power allocation $\mathbf{p}_{n}^{d*}$, the received SNR at the corresponding user $k_{n}$ can be expressed as
\begin{align}
    \gamma_{k_{n}, n}^{*} = \bigg(\frac{1}{\sigma}\sum\limits_{m \in \mathcal{M}_{n}}\sqrt{p_{m, n}^{d*}|h_{k_{n}, m, n}^d|^2} \bigg)^{2}. \label{snr}
\end{align}
Using \eqref{snr} in \eqref{Pr}, we can obtain the expression of the optimal power allocation $\mathbf{p}_{n}^{d*}$ as
\begin{align}
    p_{m, n}^{d*} = \bigg(\frac{1}{\ln 2}\bigg)^{2} \frac{\gamma_{k_{n}, n}^{*}}{\Big(1 + \gamma_{k_{n}, n}^{*}\Big)^{2}} \frac{|h_{k_{n}, m, n}^d|^2}{\sigma^2 \mu_{m}^{2}}, ~~ m \in \mathcal{M}_{n}. \label{pr}
\end{align}
Thus, we can compute the value of $p_{m, n}^{d*}$ for each $m \in \mathcal{M}_{n}$ utilizing \eqref{pr}. Substituting the value of $p_{m, n}^{d*}$ into \eqref{snr}, the expression of the optimal SNR at the user $k_{n}$ is given by
\begin{align}
    \gamma_{k_{n}, n}^{*} = \bigg(\frac{1}{\ln 2}\bigg)^{2} \frac{\gamma_{k_{n}, n}^{*}}{\big(1 + \gamma_{k_{n}, n}^{*}\big)^{2}} \Bigg(\sum\limits_{m \in \mathcal{M}_{n}} \frac{|h_{k_{n}, m, n}^d|^2}{\sigma^2 \mu_{m}}\Bigg)^{2}, \label{snr2}
\end{align}
which can be rearranged by taking the square roots at both sides and substituting the definition \eqref{Bn}. Then, we have
\begin{align}
    \gamma_{k_{n}, n}^{* 1/2} \bigg(1 + \gamma_{k_{n}, n}^{*} - \frac{1}{\ln 2} \mathcal{B}_{k_{n}, n} \bigg) = 0. \label{snr3}
\end{align}
We can easily obtain two roots $\gamma_{k_{n}, n}^{*} = 0$ and
\begin{align}
    \gamma_{k_{n}, n}^{*} = \frac{1}{\ln 2} \mathcal{B}_{k_{n}, n} - 1. \label{snr4}
\end{align}
If $\gamma_{k_{n}, n}^{*} = 0$, we can observe from \eqref{pr} that $p_{m, n}^{d*} = 0, m \in \mathcal{M}_{n}$. A non-zero power allocation requires that $\gamma_{k_{n}, n}^{*} > 0$, i.e., it must satisfy the condition that
\begin{align}
    \frac{1}{\ln 2} \mathcal{B}_{k_{n}, n} - 1 > 0. \label{snr5}
\end{align}
Moreover, the optimal power allocated to RRH $m$, which is not selected on SC $n$, is expected to be zero, i.e., $p_{m, n}^{d*} = 0$ if $m \notin \mathcal{M}_{n}$. Finally, substituting \eqref{snr4} in \eqref{pr}, along with condition \eqref{snr5} and definition \eqref{Bn}, the expression of the optimal power allocation $\mathbf{p}_{n}^{d*}$ is obtained as given in \eqref{P_r1}.
\end{IEEEproof}

For given selected user $k_{n}$ and RRH subset $\mathcal{M}_{n}$ on SC $n$, the value of $\mathcal{B}_{k_{n},n}$ in \eqref{Bn} is the same for all the RRHs in $\mathcal{M}_{n}$. However, the optimal DL power allocation in \eqref{P_r1} is varied by the channel power gain $|h_{k_n, m, n}^d|^{2}$ for each RRH $m$ in $\mathcal{M}_{n}$. This is because the DL is a MISO channel, and the transmit power of each RRH is proportional to its channel power gain \cite{MISO1}.

Note that if there is only one RRH, i.e., $M = 1$, the power allocation in \eqref{P_r1} becomes
\begin{align}
    p_{n}^{d*} = \bigg[\frac{1}{\mu \ln 2} - \frac{\sigma^2}{|h_{k_{n}, n}^{d}|^2} \bigg]^{+},
\end{align}
which follows the classical water-filling solution \cite{water-filling}.

So far, we can observe that both the power allocations $\mathbf{p}_{n}^{u*}$ and $\mathbf{p}_{n}^{d*}$ depend on the UL/DL scheduling $y_{n}$ and the RRH selection $\mathcal{M}_{n}$. Thus, when the RRH selection is given, we can obtain the optimal power allocation using the preceding solutions.

\subsubsection{Maximizing Lagrangian over $\mathbf{X}_{n}$}
For the binary variable $x_{k, m, n}$, only one $k$ but multiple $m$'s can be active on SC $n$ according to the constraint \eqref{cons4}, which implies that finding the optimal $\mathbf{X}_{n}^{*}$ is equivalent to finding a $k_{n}^{*} \in \mathcal{K}$ and a subset of RRHs $\mathcal{M}_{n}^{*} \subseteq \mathcal{M}$, i.e.,
\begin{align}
    x_{k, m, n}^{*} =
    \begin{cases}
    1,   &\mbox{if $k = k_{n}^{*}$ and $m \in \mathcal{M}_{n}^{*}$},\\
    0,   &\mbox{otherwise}.
    \end{cases}
\end{align}

The optimal solution yields to the exhaustive search which needs to search all possible RRH subsets and users. Specifically, we first list all $2^{M}$ possibilities of RRH subsets $\mathcal{M}_{n} \subseteq \mathcal{M}$. Then we choose one RRH subset $\mathcal{M}_{n}^{*}$ along with one user $k_{n}^{*}$ that maximizes the Lagrangian $\mathcal{L}_{n}$, i.e.,
\begin{align}
    \mathbf{X}_{n}^{*} = \arg \max_{\{k_{n}, \mathcal{M}_{n}\}} \mathcal{L}_{n}
\end{align}
as the optimal solution of $\mathbf{X}_{n}^{*}$.

Although finding the optimal $\mathbf{X}_{n}^{*}$ needs $\mathcal{O}\big(2^{M}K\big)$ complexity, it is not very high if the number of RRHs is small, e.g., $M=4$ or $M=8$. To avoid high complexity for large $M$, we will introduce a lower-complexity heuristic algorithm in the next section.

\subsubsection{Maximizing Lagrangian over $y_{n}$}
As a SC can be assigned to either UL or DL of a user, we can determine the optimal $y_{n}^{*}$ by choosing the larger value of $\mathcal{L}_{n}^u$ and $\mathcal{L}_{n}^{d}$, i.e.,
\begin{align}
    y_{n}^{*} =
    \begin{cases}
    1,   &\mbox{if $\mathcal{L}_{n}^{u} > \mathcal{L}_{n}^{d}$}, \\
    0,   &\mbox{otherwise}.
    \end{cases}  \label{y_n}
\end{align}

In summary, we can solve the subproblem \eqref{subproblem} optimally as follows. First, for given dual variables $\{\bm{\lambda},\bm{\mu}, \bm{\nu}\}$, fix the user assigned to SC $n$ as $k_{n} \in \mathcal{K}$. By letting $y_{n}$ as $1$ and $0$, respectively, find the corresponding optimal RRH selection via exhaustively searching all possible RRH subsets and derive the optimal power allocations $\mathbf{p}_{n}^{u*}$ and $\mathbf{p}_{n}^{d*}$ by solving problems \eqref{pa_ul} and \eqref{pa_dl}, respectively. Then find the optimal RRH selection $\mathcal{M}_{n}^{*}$ with the obtained power allocations $\mathbf{p}_{n}^{u*}$ and $\mathbf{p}_{n}^{d*}$ for the fixed user $k_{n}$ as the solution that maximizes the objective of subproblem \eqref{subproblem}.  Subsequently, we can find the optimal SC assignment $k_{n}^{*}$ on SC $n$ with the obtained optimal RRH selection and power allocation. By doing so, we can obtain two values of subproblem \eqref{subproblem} corresponding to $y_{n} = 0$ and $y_{n} = 1$, respectively, and finally we find the optimal $y_{n}^{*}$ by using \eqref{y_n}.

\subsection{Optimizing Dual Variables $\{\bm{\lambda},\bm{\mu}, \bm{\nu}\}$}

After finding the optimal $\{\mathbf{X}^{*}, \mathbf{y}^{*}, \mathbf{P}_{u}^{*},\mathbf{P}_{r}^{*}\}$, we turn to solve the dual problem which can be expressed as
\begin{align}
    \min_{\bm{\lambda} \succeq 0,\bm{\mu} \succeq 0, \bm{\nu} \succeq 0}g(\bm{\lambda},\bm{\mu}, \bm{\nu}). \label{dual problem}
\end{align}

As the dual problem is always convex according to \cite{ConvexOptimization}, we can use the ellipsoid method to simultaneously update the dual variables $\{\bm{\lambda}, \bm{\mu}, \bm{\nu}\}$ towards the optimal $\{\bm{\lambda}^{*}, \bm{\mu}^{*}, \bm{\nu}^{*}\}$ by using the subgradients obtained in the following proposition:

\begin{proposition} \label{proposition4}
The subgradients can be obtained by the definition as follows:
\begin{align}
    \Delta \lambda_{k} &= P_{k}^u -\sum\limits_{n \in \mathcal{N}} p_{k, n}^u,  ~~~~ k \in \mathcal{K}. \label{subgradient1} \\
    \Delta \mu_{m} &= P_{m}^d - \sum\limits_{n \in \mathcal{N}} p_{m, n}^d,   ~~~~ m \in \mathcal{M}. \label{subgradient2} \\
    \Delta \nu_{m} &= C_{m} - \sum\limits_{n \in \mathcal{N}}\sum\limits_{k \in \mathcal{K}} x_{k, m, n}, ~~~~ m \in \mathcal{M}. \label{subgradient3}
\end{align}
\end{proposition}

\begin{IEEEproof}
Please refer to Appendix \ref{appendicesC}.
\end{IEEEproof}

So far, we have solved Problem \eqref{problem} asymptotically optimally by the dual method through iteratively updating the dual variables. The above asymptotically optimal algorithm is described in detail in Algorithm $1$. In this algorithm, for each SC, we need to search over $2^{M}$ possible selections to find the optimal RRH subset and the complexity is $\mathcal{O}\big(2^{M}\big)$. Then we need to search over $K$ users and $2$ possible $y_{n}$ values to find the optimal SC assignment and UL/DL scheduling, whose complexity is $\mathcal{O}\big(2K\big)$. Therefore, each subproblem \eqref{subproblem} can be solved with a complexity of $\mathcal{O}\big(2^{M+1}K\big)$. The complexity of subgradient update is $\mathcal{O}\big((K + 2M)^{2}\big)$ by using the ellipsoid method \cite{ConvexOptimization}. Combining the decomposition over $N$ SCs, the overall complexity is $\mathcal{O}\big(2^{M+1}KN(K + 2M)^{2}\big)$.

\begin{algorithm}[tb]
\caption{Asymptotically optimal algorithm}
\begin{algorithmic}[1]
\STATE \textbf{Initialize} $\{\bm{\lambda, \mu, \nu}\}\geq0$.
\REPEAT
    \FOR{each $n \in \mathcal{N}$}
        \FOR{each $y_{n} \in \{0, 1\}$}
            \FOR{each user association $k_{n}$}
                \STATE Derive power allocations by solving problem \eqref{pa_ul} if $y_n = 1$, otherwise solving problem \eqref{pa_dl}, for $2^{M}$ possible RRH selections.
                \STATE Find the optimal RRH selection $\mathcal{M}_{n}^{*}$ and the corresponding power allocations $\mathbf{p}_{n}^{u*}$ and $\mathbf{p}_{n}^{d*}$ that maximizes \eqref{subproblem}.
            \ENDFOR
            \STATE Find the optimal user association $k_{n}^{*}$ that maximizes \eqref{subproblem} with $\mathcal{M}_{n}^{*}$, $\mathbf{p}_{n}^{u*}$ and $\mathbf{p}_{n}^{d*}$ obtained in Line 7.
        \ENDFOR
        \STATE Choose the value of $y_{n}^{*}$ that maximizes \eqref{subproblem}.
    \ENDFOR
    \STATE Update $\{\bm{\lambda, \mu, \nu}\}$ by the ellipsoid method using the subgradients defined in \eqref{subgradient1}-\eqref{subgradient3}.
\UNTIL{$\{\bm{\lambda, \mu, \nu}\}$ converge.}
\end{algorithmic}
\end{algorithm}

\begin{algorithm}[tb]
\caption{Heuristic algorithm}
\begin{algorithmic}[1]
\STATE \textbf{Initialize} $\{\bm{\lambda, \mu, \nu}\}\geq0$.
\REPEAT
    \FOR{each $n \in \mathcal{N}$}
        \FOR{each $y_{n} \in \{0, 1\}$}
            \FOR{each user association $k_{n}$}
                \STATE \textbf{Initialize} $\mathcal{M}_{n} = \emptyset, V^{1} = 0$.
                \IF{$y_{n} = 0$}
                    \STATE Sort$(h_{k, m, n}^d,$ descend$)$.
                \ELSE
                    \STATE Sort$(h_{k, m, n}^u,$ descend$)$.
                \ENDIF

                \FOR{each $l = 1, \ldots, M$}
                    \STATE Select RRH $m_{l}$ according to the sorted channels.
                    \IF{$m_{l}$ satisfies condition \eqref{selection}}
                        \STATE Update selected RRH set as $\mathcal{M}_{n} = \mathcal{M}_{n} \cup \{m_{l}\}$.
                        \STATE Update objective $V^{l + 1}$ according to \eqref{objective}.
                    \ELSE
                        \STATE Remain $\mathcal{M}_{n}$ unchanged and $V^{l + 1} = V^{l}$.
                    \ENDIF
                \ENDFOR
                \STATE Derive power allocations by solving problem \eqref{pa_ul} if $y_n = 1$, otherwise solving problem \eqref{pa_dl}, for the obtained RRH selections.
            \ENDFOR
            \STATE Find the optimal user association $k_{n}^{*}$ that maximizes \eqref{subproblem}.
        \ENDFOR
        \STATE Choose the value of $y_{n}^{*}$ that maximizes \eqref{subproblem}.
    \ENDFOR
    \STATE Update $\{\bm{\lambda, \mu, \nu}\}$ by the ellipsoid method using the subgradients defined in \eqref{subgradient1}-\eqref{subgradient3}.
\UNTIL{$\{\bm{\lambda, \mu, \nu}\}$ converge.}

\end{algorithmic}
\end{algorithm}

\section{Heuristic Solution}

As the proposed asymptotically optimal algorithm obtains the optimal RRH selection via an exhaustive search, the complexity becomes prohibitive if the number of RRHs $M$ becomes large. Hence, we turn to propose a practical heuristic solution to simplify the RRH selection in this section, which significantly reduces the complexity. Note that this section only focuses on RRH selection and the rest of optimization is the same as the proposed asymptotically optimal algorithm.

Given SC assignment, UL/DL scheduling and power allocation on SC $n$, the objective of subproblem \eqref{subproblem} can be expressed as $V(\mathcal{M}_{n})$, a function of the RRH selection $\mathcal{M}_{n}$. Since the transmit power allocations are fixed, each user tends to select the RRHs with better channels. Moreover, to utilize the resource efficiently, we require that each selected RRH $m \in \mathcal{M} \setminus \mathcal{M}_{n}$ increases the objective value of subproblem \eqref{subproblem}, i.e.,
\begin{align}
    V\big(\mathcal{M}_{n} \cup \{m\}\big) > V\big(\mathcal{M}_{n}\big). \label{selection}
\end{align}

Inspired by these, we solve the RRH selection in $M$ iterations as follows. Let $V^{l}$ denote the objective value at iteration $l$. We first initialize that $\mathcal{M}_{n} = \emptyset$ and $V^{1} = 0$. We sort the channel power gains on SC $n$ in the descending order. To be specific, sort the values of $h_{k, m, n}^u$ when $y_{n} = 1$, and sort the values of $h_{k, m, n}^d$ when $y_{n} = 0$, so that the channel power gain indicates the priority of the corresponding RRH to be selected. Then, we select the RRH $m_{l}$ with the $l^{th}$ largest channel power gain at each iteration $l = 1, \cdots , M$. Upon the selection of each RRH $m_{l}$, the condition \eqref{selection} is checked. RRH $m_{l}$ is added to the set of selected RRHs $\mathcal{M}_{n}$ if it increases the objective value $V^{l}$, then the set is updated as $\mathcal{M}_{n} = \mathcal{M}_{n} \cup \{m_{l}\}$ and the objective value is given by
\begin{align}
    V^{l + 1} = V\big(\mathcal{M}_{n} \cup \{m_{l}\} \big). \label{objective}
\end{align}
After $M$ iterations, we can obtain the final suboptimal RRH selection $\mathcal{M}_{n}$ on SC $n$ and the corresponding power allocations. The above heuristic algorithm is described in detail in Algorithm $2$.

In this algorithm, the computational complexity of sorting the channels is $\mathcal{O}\big(M\log_{2}(M)\big)$, and the complexity of $M$ iterations is $\mathcal{O}\big(M\big)$. Note that they are independent. Therefore, the overall complexity of solving Problem \eqref{problem} is reduced to $\mathcal{O}\big(M\log_{2}(M)2KN(K + 2M)^{2}\big)$, which is much lower than that of the proposed asymptotically optimal algorithm when the value of $M$ is large.

\section{TDD Case}
The proposed framework can be extended to the TDD mode for UL and DL transmissions, where all SCs are assigned to either UL or DL in each time slot. Let $y$ indicate the UL/DL scheduling on all SCs in a given time slot, i.e.,
\begin{align}
    y =
    \begin{cases}
    1,   &\mbox{if all SCs are assigned to UL},\\
    0,   &\mbox{if all SCs are assigned to DL}.
    \end{cases}
\end{align}
Then, the problem can be formulated as follow:
\begin{subequations}
\begin{align}
\max_{\{\mathbf{X}, y, \mathbf{P}_{u},\mathbf{P}_{r}\}}~&R_{total}=y \sum\limits_{n \in \mathcal{N}}\sum\limits_{k \in \mathcal{K}}w_{k} R_{k,n}^{u} \nonumber \\
&~~~~~~~~~~+ (1 - y)\sum\limits_{n \in \mathcal{N}}\sum\limits_{k \in \mathcal{K}} R_{k,n}^{d} \label{problem_tdd} \\
{\rm s.t.}~&\eqref{cons1}-\eqref{cons5},    \nonumber \\
&y \in \{0, 1\}.
\end{align}
\end{subequations}

The analysis and solutions of Problem \eqref{problem_tdd} are similar to the flexible FDD case. That is, by letting $y = 1$ and $y = 0$, respectively, we can find the corresponding optimal $\{\mathbf{X}, \mathbf{P}_{u},\mathbf{P}_{r}\}$ by carrying out Algorithm $1$. Then we choose one case of $y$ and the corresponding $\{\mathbf{X}, \mathbf{P}_{u},\mathbf{P}_{r}\}$ that has a larger objective value. Similarly, Algorithm $2$ is also applicable to Problem \eqref{problem_tdd}.

\section{Numerical Results}

\begin{figure}[tb]
    \begin{centering}
        \includegraphics[scale=0.64]{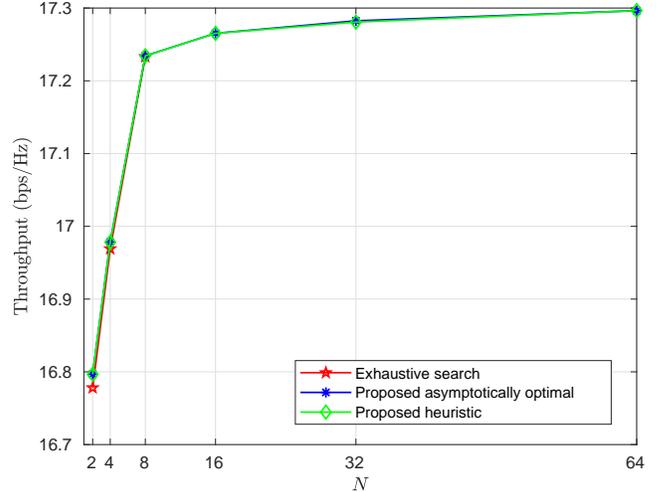}
        \vspace{-0.1cm}
         \caption{Throughput versus the number of SCs $N$, where $K = 2$ and $M = 4$.}\label{fig:gap}
        \end{centering}
    \vspace{-0.1cm}
\end{figure}

In this section, numerical results are provided to evaluate the performance of the proposed schemes. In the simulation, the users and RRHs are uniformly distributed within a square area with a length of $500$ meters. We set the carrier center frequency as $2$ GHz and the total bandwidth as $10$ MHz, which is divided into $N = 64$ SCs using OFDMA. The resolution of uniform scalar quantization at the RRHs is set as $\beta = 10$. The maximum number of accessible SCs is assumed to be the same for all the RRHs, i.e., $C_{m} = C, \forall m \in \mathcal{M}$. We assume that the transmit power of users and RRHs are identical, respectively, i.e., $P_k^u = P_u, \forall k \in \mathcal{K}$, and $P_m^d = P_r, \forall m \in \mathcal{M}$. The small-scale fading is assumed to be Rayleigh, and the path loss exponent is $3$. The noise spectral density is set as $-174$ dBm/Hz. Without loss of generality, we consider the sum rate maximization in Problem \eqref{problem}, i.e, the UL user rate weights $w_{k} = 1, \forall k \in \mathcal{K}$.

First, we compare the two proposed algorithms with the globally optimal solution by exhaustive search in Fig. \ref{fig:gap}. We consider a small-size network with $M = 4, K = 2$, and we set the number of accessible SCs $C = N$. Note that we do not plot the curve of the exhaustive search algorithm when $N \geq 16$ due to the prohibitive complexity. Fig. \ref{fig:gap} shows that the duality gap indeed exists for the dual-based method, and the obtained results by the dual-based method are the upper-bounds for the original throughput maximization problem. However, the dual gap becomes zero when the practical number of SCs $N > 8$. This validates the proposed algorithms.

\begin{table}[t]
\caption{Running time comparison}
\centering
\begin{tabular}{|c|c|c|c|}
  \hline
  \multirow{2}*{Algorithm} & \multicolumn{3}{c|}{Running time (s)}  \\ \cline{2-4}
                            &  $N = 8$  &   $N = 16$  & $N = 64$ \\
  \hline
  Optimal via exhaustive search & 38976.228 & - & - \\ \hline
  Proposed asymptotically optimal & 29.129 & 51.196 & 232.329 \\ \hline
  Proposed heuristic & 10.579 & 18.789 & 78.755 \\

  \hline
\end{tabular}
\label{table:time}
\end{table}

The running time of these three algorithms is listed in Table \ref{table:time}, which are evaluated under the hardware environment with 2.20 GHz CPU and 64 GB memory. From Table \ref{table:time}, we can observe that the running time of the exhaustive search algorithm is much more than the proposed algorithms when $N = 8$. It is also observed that the proposed heuristic algorithm can achieve a close-to-optimal performance as the proposed asymptotically optimal algorithm with much less time. The running time of the proposed asymptotically optimal algorithm is near twice as much as that of the proposed heuristic algorithm for all $N$, which is consistent with the comparison of their complexity.

Next, we consider the following benchmark schemes for the goal of performance comparison.
\begin{itemize}
  \item  \textbf{Equal power allocation (EPA)}. In this scheme, the transmit power allocated at all the user and RRHs are set as $p_{k, n}^u = P_{u}/ N, \forall k \in \mathcal{K}$ and $p_{m, n}^d = P_{r} / C, \forall m \in \mathcal{M}$, respectively. Therefore, Problem \eqref{problem} is reduced to find the optimal UL/DL scheduling, RRH selection and SC assignment. The proposed asymptotically optimal algorithm can also solve this problem, but the subgradient update \eqref{subgradient1} and \eqref{subgradient2} are eliminated and the complexity decreases to $\mathcal{O}\big(2^{M+1}KNM^{2}\big)$.
  \item  \textbf{Average SC assignment (ASA)}. The SC assignment in this scheme is fixed, i.e., the total $N$ SCs are averagely assigned to $K$ users and thus $\lfloor N / K \rfloor$ SCs are assigned to each user. Thus, the problem is simplified as a joint UL/DL scheduling, RRH selection and power allocation problem. The overall complexity of the proposed asymptotically optimal algorithm for this scheme is reduced to $\mathcal{O}\big(2^{M+1}N(K + 2M)^{2}\big)$.
  \item  \textbf{Nearest RRH selection (NRS)}. Each user selects its corresponding nearest RRH on all assigned SCs in this scheme. Thus, the problem is simplified as a joint UL/DL scheduling, SC assignment and power allocation problem. The overall complexity of the proposed asymptotically optimal algorithm for this scheme is thus reduced to $\mathcal{O}\big(2KN(K + 2M)^{2}\big)$.
\end{itemize}

\begin{figure}[t]
\begin{centering}
\includegraphics[scale=0.64]{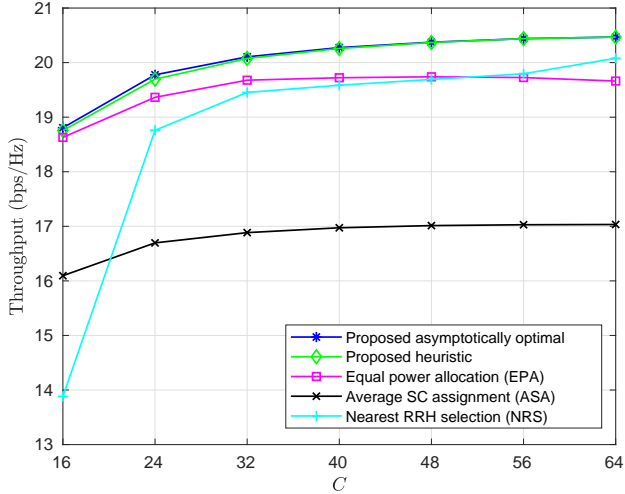}
\vspace{-0.1cm}
 \caption{Throughput versus the number of accessible SCs $C$, where $K = 8$ and $M = 4$.}\label{fig:SC}
\end{centering}
\vspace{-0.1cm}
\end{figure}

Fig. \ref{fig:SC} demonstrates the throughput performance versus the number of accessible SCs $C$, where the number of users $K = 8$, the number of RRHs $M = 4$, the transmit power $P_{u} = 23$ dBm and $P_{r} = 30$ dBm are fixed. From Fig. \ref{fig:SC}, we can observe that the performance of the proposed heuristic algorithm is close to the proposed asymptotically optimal algorithm. The proposed algorithms outperform the three benchmark schemes, indicating that jointly optimizing RRH selection, SC assignment, power allocation and UL/DL scheduling is necessary. When $C = 64$, all the RRHs are selected on each SC in all schemes except the NRS scheme. Nonetheless, the NRS scheme can still obtain performance gain because the best RRH is selected on each SC. We can observe that when $C = 16$, the NRS scheme is the worst one, but it outperforms the ASA scheme and the EPA scheme as $C$ increases. Besides, it is also observed that the throughput of the EPA scheme increases first and decreases subsequently, this is because the loss of equal power allocation exceeds the gain of increasing the number of accessible SCs, showing the imperfection of the equal power allocation. We set $C = N/2 = 32$ in the subsequent simulations unless specified.

\begin{figure}[t]
\begin{centering}
\includegraphics[scale=0.64]{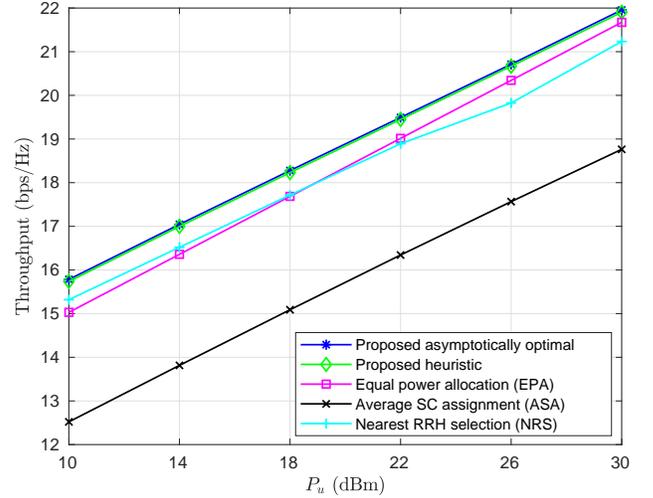}
\vspace{-0.1cm}
 \caption{Throughput versus the transmit power of users $P_{u}$, where $K = 8$ and $M = 4$.}\label{fig:power}
\end{centering}
\vspace{-0.1cm}
\end{figure}

Fig. \ref{fig:power} shows the throughput performance versus the transmit power of users $P_{u}$. In this case, we set $M = 4$ and $K = 8$. We also let the transmit power of RRHs $P_{r} = (P_{u} + 7)$ dBm. From Fig. \ref{fig:power}, we can first observe that the throughput increases linearly with the transmit power. It is also observed that the performance of the proposed heuristic algorithm is close to the proposed asymptotically optimal algorithm for all $P_u$. The performance gap between the EPA scheme and the proposed algorithms becomes smaller as $P_u$ increases. This is because the benefit of power allocation is limited when the SNR is sufficiently high. The performance of the ASA scheme is the worst, showing the performance gain of the optimal SC assignment.

\begin{figure}[t]
\begin{centering}
\includegraphics[scale=0.64]{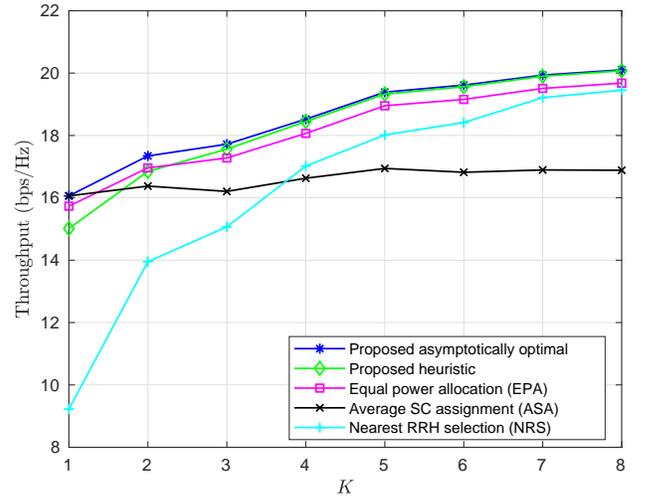}
\vspace{-0.1cm}
 \caption{Throughput versus the number of users $K$, where $M = 4$.}\label{fig:user}
\end{centering}
\vspace{-0.1cm}
\end{figure}

Fig. \ref{fig:user} illustrates the throughput performance versus the number of users $K$, where the number of RRHs $M = 4$, the transmit power $P_{u} = 23$ dBm and $P_{r} = 30$ dBm are fixed. It can be observed that the proposed asymptotically optimal algorithm has the best performance for all $K$. When the number of users $K$ is small, the performance of the proposed heuristic algorithm is not ideal. However, the performance gap between the proposed heuristic algorithm and the proposed asymptotically optimal algorithm becomes smaller as $K$ increases. This is because the RRH selection is not optimal in the proposed heuristic algorithm, but the RRHs are more likely to be selected by the users that can achieve higher performance as the number of users increases. This indicates that the proposed heuristic algorithm performs well under the condition that the number of users is large enough, which is generally satisfied in practice. When $K = 1$, the performance of the ASA scheme is equal to the proposed asymptotically optimal algorithm, this is because all the SCs will be assigned to the single user in these two schemes. Since the ASA scheme performs the equal assignment of SCs to users, it benefits less from the increase of $K$ and the throughput remains almost the same. The throughput of the NRS scheme increases significantly with $K$ although the performance is the worst when $K$ is small.

\begin{figure}[t]
\begin{centering}
\includegraphics[scale=0.64]{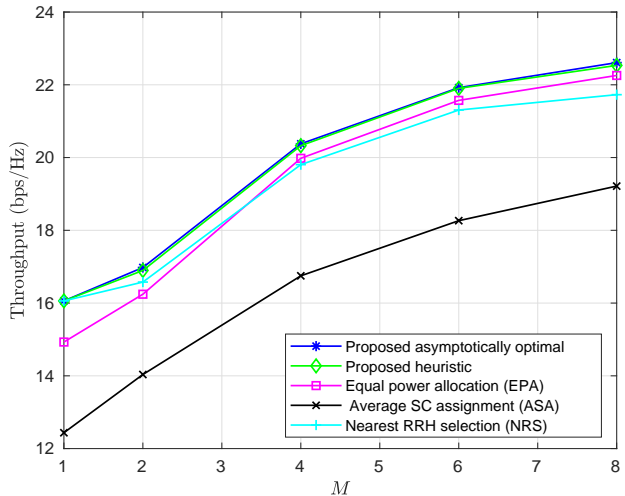}
\vspace{-0.1cm}
 \caption{Throughput versus the number of RRHs $M$, where $K = 8$.}\label{fig:RRH}
\end{centering}
\vspace{-0.1cm}
\end{figure}

Fig. \ref{fig:RRH} illustrates the throughput performance versus the number of RRHs $M$, where the number of users $K = 8$, the transmit power $P_{u} = 23$ dBm and $P_{r} = 30$ dBm are fixed. Specifically, we set $C = N$ for $M = 1$ to make sure that all the SCs are utilized in this scenario, while $C = N/2$ for others. From Fig. \ref{fig:RRH}, it is first observed that the throughput achieved by the proposed heuristic algorithm is nearly equal to the proposed asymptotically optimal algorithm for all $M$, which further validates that the proposed heuristic algorithm has a close-to-optimal performance. We can also observe that the throughput of the NRS scheme is the same as the proposed asymptotically optimal algorithm when $M = 1$. In this scenario, all the users will select the single RRH, so the NRS scheme will obtain the optimal RRH selection. As the number of RRHs $M$ increases, the performance gap between the NRS scheme and the proposed algorithms becomes larger, showing the benefit of CoMP by multiple RRHs.

From Figs. \ref{fig:SC}$-$\ref{fig:RRH}, we can conclude that the performance of the proposed heuristic algorithm is close to the proposed asymptotically optimal algorithm. We next simulate large networks to test the performance of the proposed heuristic algorithm against the benchmark schemes. For a fair comparison, the benchmarks here are developed based on the proposed heuristic algorithm.

\begin{figure}[t]
\begin{centering}
\includegraphics[scale=0.64]{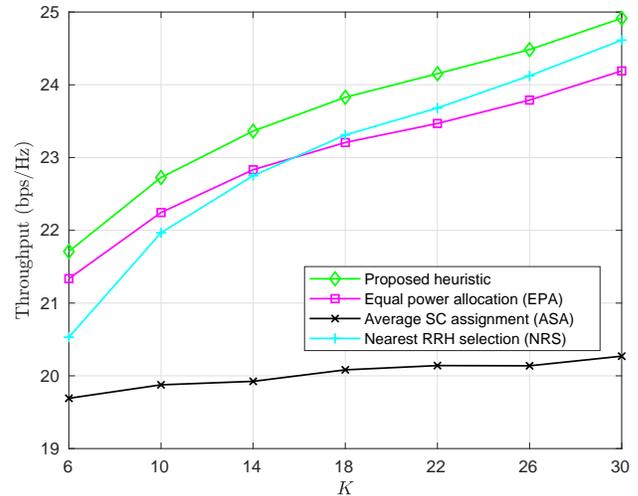}
\vspace{-0.1cm}
 \caption{Throughput versus the number of users $K$, where $M = 10$.}\label{fig:user_large}
\end{centering}
\vspace{-0.1cm}
\end{figure}

Fig. \ref{fig:user_large} plots the throughput performance versus the number of users $K$, where $M = 10$. We observe that the performance comparison among the schemes is almost consistent with Fig. \ref{fig:user}. As $K$ increases, the proposed heuristic algorithm achieves much higher throughput than all benchmark schemes except the NRS scheme. This is because the RRHs are more likely to be selected by the users that can achieve higher throughput as the number of users increases. Besides, the throughput of the ASA scheme increases slowly with $K$, since it cannot make much of the diversity gain provided by larger $K$.

\begin{figure}[t]
\begin{centering}
\includegraphics[scale=0.64]{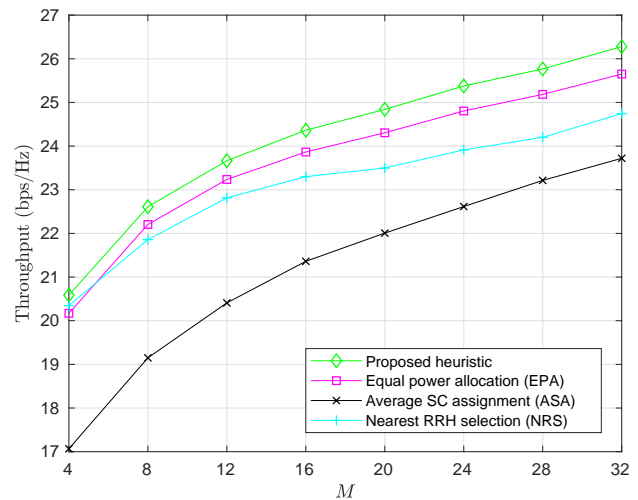}
\vspace{-0.1cm}
 \caption{Throughput versus the number of RRHs $M$, where $K = 10$.}\label{fig:RRH_large}
\end{centering}
\vspace{-0.1cm}
\end{figure}

Fig. \ref{fig:RRH_large} shows the throughput performance versus the number of RRH $M$, where $K = 10$. Again, the performance comparison among the schemes is similar to Fig. \ref{fig:RRH}. Since the NRS scheme cannot exploit the benefit of CoMP by multiple RRHs, the throughput increases slower than other schemes and it tends to be the worst one as $M$ increases.

\begin{figure}[t]
\begin{centering}
\includegraphics[scale=0.64]{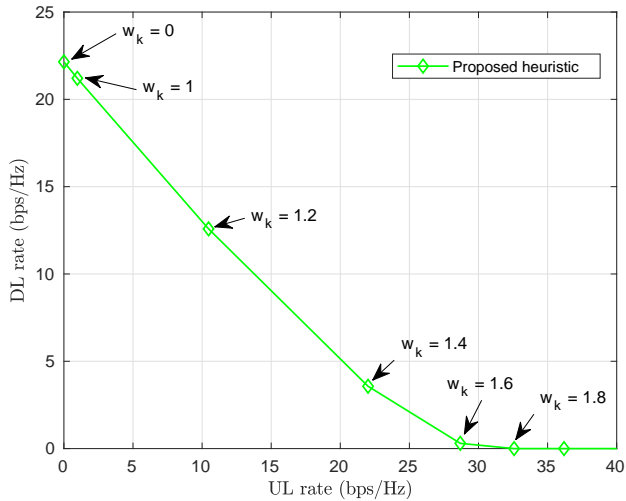}
\vspace{-0.1cm}
\caption{DL rate versus UL rate, where $K = 10$ and $M = 10$.}\label{fig:region}
\end{centering}
\vspace{-0.1cm}
\end{figure}

Fig. \ref{fig:region} illustrates the achievable rate region of the proposed heuristic algorithm, where $K = 10$, $M = 10$, $P_u = 23$ dBm and $P_r = 30$ dBm. The UL rate is determined by the UL user rate weights $w_k, \forall k \in \mathcal{K}$. Therefore, Fig. \ref{fig:region} shows the impact of $w_k$ for the whole resource allocation. We observe that when $w_k = 1, \forall k \in \mathcal{K}$, which is actually the case of sum rate maximization, the DL rate is much larger than the UL rate since the UL power level is less than that of the DL and more SCs are assigned to the DL. It is also observed that when the UL rate is increasing, the DL rate finally converges to zero. This is because when the weights $w_k, \forall k \in \mathcal{K}$ are large enough, i.e., $w_k > 1.6, \forall k \in \mathcal{K}$, all the SCs are scheduled to perform UL transmission.

\begin{figure}[t]
\begin{centering}
\includegraphics[scale=0.64]{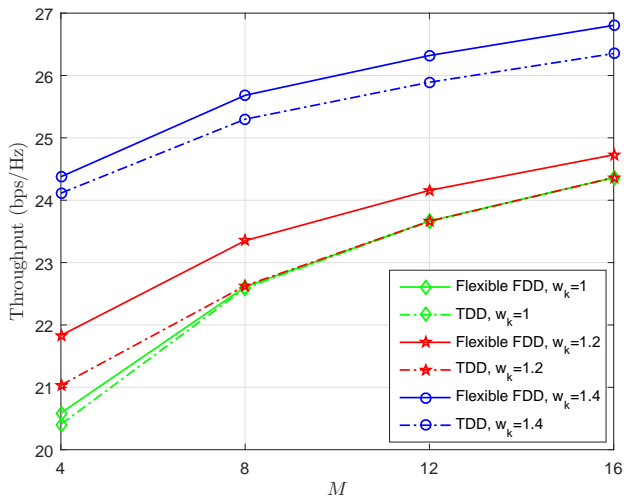}
\vspace{-0.1cm}
 \caption{Throughput versus the number of RRHs $M$, where $K = 10$.}\label{fig:FDD_TDD}
\end{centering}
\vspace{-0.1cm}
\end{figure}

Finally, we compare the throughput performance between the flexible FDD case and the TDD case in Fig. \ref{fig:FDD_TDD}. We plot the throughput versus the number of RRHs $M$ under different values of $w_k$. It shows that the throughput of the TDD case is lower than that of the flexible FDD case, which is because the flexible FDD case is more flexible. In the case of $w_k = 1$, the performance gap between the flexible FDD case and TDD case is very small. This is because the weighted UL rate is much smaller than the DL rate, and the flexibility of the flexible FDD case does not provide much benefit. When $w_k = 1.2$ and $w_k = 1.4$, the weighted UL rate is comparable to the DL rate, thus the flexible FDD case provides higher throughput than the TDD case. We also observe that the performance of the TDD case is almost the same for $w_k = 1$ and $w_k = 1.2$ when $M$ is large. 
These show that the flexible FDD case is more sensitive to the rate than the TDD case, thus the flexible FDD case achieves better performance.

\section{Conclusion}

In this paper, we investigated the joint UL and DL resource allocation in an OFDMA-based CRAN. In particular, we considered CoMP on each SC together with UL/DL decoupling. The optimization problem of joint UL/DL scheduling, SC assignment, RRH selection and power allocation for maximizing the system throughput was studied. We proposed an efficient algorithm based on the Lagrange duality method to solve this non-convex and NP-hard problem asymptotically optimally. Moreover, we proposed a heuristic algorithm which has a close-to-optimal performance with much lower complexity. Numerical results showed that the proposed algorithms can achieve higher throughput compared to benchmark schemes.

\appendices
\section{Proof of NP-hardness} \label{appendicesA}
The subproblem of joint scheduling of SC assignment and RRH selection to maximize the throughput is actually a maximum weight clique problem \cite{graph1,graph2}, which is NP-hard. We can use $\mathcal{A}$ to denote the set of all possible associations between users, RRHs and SCs, i.e., $\mathcal{A} = \mathcal{K} \times \mathcal{M} \times \mathcal{N}$. The power-set of $\mathcal{A}$ can be denoted by $\mathcal{P}(\mathcal{A})$. Note that $\mathcal{P}(\mathcal{A})$ is the set of all possible schedules of SC assignment and RRH selection regardless of the constraints. Let $\mathcal{F}$ be the set of all feasible schedules satisfying the constraints, and it is obvious that $\mathcal{F} \subset \mathcal{P}(\mathcal{A})$. Let $\mathbf{S} = \{s_1, \cdots, s_{|\mathbf{S}|}\} \in \mathcal{F}$ be any feasible schedule where $s_i \in \mathcal{A}, \forall i \leq |\mathbf{S}|$. Define $f: \mathcal{A} \rightarrow \mathbb{R}$ as a function mapping from each individual association $s_i$ to the achievable rate. The subproblem can be formulated as follows:
\begin{align}
    \max ~&\sum_{i = 1}^{|\mathbf{S}|} f(s_i) \label{scheduling} \\
    {\rm s.t.} ~~&~\mathbf{S} \in \mathcal{F}. \nonumber
\end{align}
We can build the corresponding scheduling graph $\mathcal{G}(\mathcal{V}, \mathcal{E})$ where each vertex $v \in \mathcal{V}$ is an association between users, RRHs and SCs, and the distinct vertices are connected by an edge in $\mathcal{E}$ if the constraints are satisfied. Define $\mathcal{C}$ as the set of all possible cliques with degree $Z_{total}$. Then, the problem \eqref{scheduling} can be written as a maximum weight clique problem in the following:
\begin{align}
    \mathbf{S}^* &= \argmax\limits_{\mathbf{S} \in \mathcal{F}} \sum_{i = 1}^{|\mathbf{S}|} f(s_i) \nonumber \\
    &= \argmax\limits_{\mathbf{C} \in \mathcal{C}} \sum_{i = 1}^{|\mathbf{C}|} w(v_i),
\end{align}
where $\mathbf{C} = \{v_1, \cdots, v_{|\mathbf{C}|}\} \in \mathcal{C}$ is a clique in the scheduling graph, and $w(v_i)$ is the weight of each vertex $v_i, \forall 1 \leq i \leq |\mathbf{C}|$. The optimal solution of the subproblem is the maximum weight clique of certain degree $Z_{total}$ in the scheduling graph where the weight of each vertex $v_i \in \mathcal{V}$ is defined as the achievable rate of its corresponding association:
\begin{align}
    w(v_i) = f(s_i).
\end{align}

\section{Proof of Proposition \ref{proposition1}} \label{appendicesB}
We prove that the time-sharing condition is satisfied in our studied problem:

Let $\{\mathbf{P}_{u, a}(\mathbf{X}, \mathbf{y})^*, \mathbf{P}_{r, a}(\mathbf{X}, \mathbf{y})^*\}$ be the optimal solution of Problem \eqref{problem} with power constraints $\{\mathbf{p}_{K, a}, \mathbf{p}_{M, a}\}$, where $\mathbf{p}_{K, a} = [P_{1, a}^u, \cdots, P_{K, a}^u]^T$ and $\mathbf{p}_{M, a} = [P_{1, a}^d, \cdots, P_{M, a}^d]^T$. Here, we have $p_{k, n, a}^{u*} > 0$ if $\sum_{m \in \mathcal{M}} x_{k, m, n} > 0$ and $y_n = 1$, otherwise $p_{k, n, a}^{u*} = 0$. Besides, $p_{m, n, a}^{d*} > 0$ if $\sum_{k \in \mathcal{K}} x_{k, m, n} > 0$ and $y_n = 0$, otherwise $p_{m, n, a}^{d*} = 0$. Similarly, let $\{\mathbf{P}_{u, b}(\mathbf{X}, \mathbf{y})^*, \mathbf{P}_{r, b}(\mathbf{X}, \mathbf{y})^*\}$ be the optimal solution of Problem \eqref{problem} with power constraints $\{\mathbf{p}_{K, b}, \mathbf{p}_{M, b}\}$, where $\mathbf{p}_{K, b} = [P_{1, b}^u, \cdots, P_{K, b}^u]^T$ and $\mathbf{p}_{M, b} = [P_{1, b}^d, \cdots, P_{M, b}^d]^T$. Let the achievable rate in the two cases be $R_{total, a}^*$ and $R_{total, b}^*$, respectively. To prove the time-sharing condition, we need to construct $\{\mathbf{P}_{u, c}(\mathbf{X}, \mathbf{y}), \mathbf{P}_{r, c}(\mathbf{X}, \mathbf{y})\}$ such that
    \begin{align}
        R_{total, c} \geq \delta R_{total, a}^* + (1 - \delta) R_{total, b}^*, \label{ts_cons1}
    \end{align}
    \begin{align}
        \sum_{n \in \mathcal{N}} p_{k, n, c}^u \leq \delta P_{k, a}^u + (1 - \delta) P_{k, b}^u, \forall k \in \mathcal{K}, \label{ts_cons2}
    \end{align}
    \begin{align}
        \sum_{n \in \mathcal{N}} p_{m, n, c}^d \leq \delta P_{m, a}^d + (1 - \delta) P_{m, b}^d, \forall m \in \mathcal{M}, \label{ts_cons3}
    \end{align}
    hold for any $0 \leq \delta \leq 1$. In practical OFDMA systems, the total bandwidth is divided into a set of SCs. As the number of SCs $N$ increases, the bandwidth of each SC becomes smaller and the channel gain within each SC approaches a constant value. When $N \rightarrow \infty$, the channel gains of adjacent SCs are approximately the same, which achieves frequency sharing. As a result, the original bandwidth of each SC can be divided into two portions that have the same channel gain, one is $\delta$ and the other is $(1 - \delta)$. Therefore, $\{\mathbf{P}_{u, c}(\mathbf{X}, \mathbf{y}), \mathbf{P}_{r, c}(\mathbf{X}, \mathbf{y})\}$ can be constructed by interleaving $\{\mathbf{P}_{u, a}(\mathbf{X}, \mathbf{y})^*, \mathbf{P}_{r, a}(\mathbf{X}, \mathbf{y})^*\}$ and $\{\mathbf{P}_{u, b}(\mathbf{X}, \mathbf{y})^*, \mathbf{P}_{r, b}(\mathbf{X}, \mathbf{y})^*\}$ in the frequency domain with a proportionality $\delta$, i.e., $\mathbf{P}_{u, c}(\mathbf{X}, \mathbf{y}) = \delta \mathbf{P}_{u, a}(\mathbf{X}, \mathbf{y})^* + (1 - \delta) \mathbf{P}_{u, b}(\mathbf{X}, \mathbf{y})^*$ and $\mathbf{P}_{r, c}(\mathbf{X}, \mathbf{y}) = \delta \mathbf{P}_{r, a}(\mathbf{X}, \mathbf{y})^* + (1 - \delta) \mathbf{P}_{r, b}(\mathbf{X}, \mathbf{y})^*$. The optimal $\{\mathbf{P}_{u, a}(\mathbf{X}, \mathbf{y})^*, \mathbf{P}_{r, a}(\mathbf{X}, \mathbf{y})^*\}$ and $\{\mathbf{P}_{u, b}(\mathbf{X}, \mathbf{y})^*, \mathbf{P}_{r, b}(\mathbf{X}, \mathbf{y})^*\}$ are constant vectors due to the channel flatness over neighbouring SCs. Clearly, the conditions \eqref{ts_cons2}-\eqref{ts_cons3} can always be satisfied for any $0 \leq \delta \leq 1$. Furthermore, the solution $\{\mathbf{P}_{u, c}(\mathbf{X}, \mathbf{y}), \mathbf{P}_{r, c}(\mathbf{X}, \mathbf{y})\}$ also achieves a rate $\delta R_{total, a}^* + (1 - \delta) R_{total, b}^*$, which satisfies \eqref{ts_cons1}. Therefore, the time-sharing condition holds for Problem \eqref{problem} as $N \rightarrow \infty$.

\section{Proof of Proposition \ref{proposition4}} \label{appendicesC}
By the definition of the Lagrange dual function $g(\bm{\lambda},\bm{\mu}, \bm{\nu})$ in \eqref{lagrangian function}-\eqref{dual function}, we have
\begin{align}
    g(\bm{\lambda}',\bm{\mu}', \bm{\nu}') \geq &\sum\limits_{n \in \mathcal{N}}\sum\limits_{k \in \mathcal{K}}\Big[w_{k} y_{n} R_{k,n}^{u} + (1 - y_{n}) R_{k,n}^{d}\Big] \nonumber \\
    &- \sum\limits_{k \in \mathcal{K}} \lambda_{k}' \bigg(\sum\limits_{n \in \mathcal{N}} p_{k, n}^u - P_{k}^u\bigg) \nonumber \\
    &- \sum\limits_{m \in \mathcal{M}} \mu_{m}' \bigg(\sum\limits_{n \in \mathcal{N}} p_{m, n}^d - P_{m}^d\bigg) \nonumber \\
    & - \sum\limits_{m \in \mathcal{M}} \nu_{m}' \bigg(\sum\limits_{n \in \mathcal{N}}\sum\limits_{k \in \mathcal{K}} x_{k, m, n} - C_{m}\bigg) \nonumber \\
    = & g(\bm{\lambda},\bm{\mu}, \bm{\nu}) + \sum\limits_{k \in \mathcal{K}} (\lambda_{k}' - \lambda_{k}) \bigg(P_{k}^u - \sum\limits_{n \in \mathcal{N}} p_{k, n}^u\bigg) \nonumber \\
    &+ \sum\limits_{m \in \mathcal{M}} (\mu_{m}' - \mu_{m}) \bigg(P_{m}^d - \sum\limits_{n \in \mathcal{N}} p_{m, n}^d\bigg) \nonumber \\
    &+ \sum\limits_{m \in \mathcal{M}} (\nu_{m}' - \nu_{m}) \bigg(C_{m} - \sum\limits_{n \in \mathcal{N}}\sum\limits_{k \in \mathcal{K}} x_{k, m, n}\bigg).
\end{align}
According to the definition of subgradient, we can obtain the subgradients in equations \eqref{subgradient1}-\eqref{subgradient3}.

\bibliography{references}

\bibliographystyle{IEEEtran}

%
%
%
\end{document}